\documentclass[
 reprint,
 aps,
 prl,
]{revtex4-1}

\usepackage{chngcntr}
\usepackage{graphicx}
\usepackage{dcolumn}
\usepackage{bm}
\usepackage[separate-uncertainty=true]{siunitx}

\newcommand{\td}{t}
\newcommand{\NA}{N}
\newcommand{\Rsc}{R_{\text{sc}}}
\newcommand{\np}{n_{\text{p}}}
\newcommand{\topt}{t_{\text{opt}}}
\newcommand{\Nmax}{N_{\text{max}}}
\newcommand{\sigN}{\sigma_N}
\newcommand{\sigNA}{\sigma_N}
\newcommand{\avarA}{\sigNA^2}
\newcommand{\avar}{\sigN^2}
\newcommand{\sigf}{\alpha}

\newcommand{\ket}[1]{|#1\rangle}

\newcommand{\cmean}{\bar{c}}
\newcommand{\sigc}{\sigma_c}

\newcommand{\dt}{\delta t}

\begin{document}

\title{Accurate Atom Counting in Mesoscopic Ensembles}

\author{D. B. Hume}
\author{I. Stroescu}
\author{M. Joos}
\author{W. Muessel}
\author{H. Strobel}
\author{M. K. Oberthaler}
\email{atomcounting@matterwave.de}
\affiliation{Kirchhoff-Institute for Physics, University of Heidelberg, INF 227, 69120 Heidelberg, Germany}

\date{\today}

\begin{abstract}
Many cold atom experiments rely on precise atom number detection, especially in the context of quantum-enhanced metrology where effects at the single particle level are important. Here, we investigate the limits of atom number counting via resonant fluorescence detection for mesoscopic samples of trapped atoms. We characterize the precision of these fluorescence measurements beginning from the single-atom level up to more than one thousand.  By investigating the primary noise sources, we obtain single-atom resolution for atom numbers as high as 1200.  This capability is an essential prerequisite for future experiments with highly entangled states of mesoscopic atomic ensembles.
\end{abstract}

\maketitle

The Heisenberg uncertainty principle sets a fundamental limit, $\Delta\phi= 1/N$, on the precision at which one can determine an interferometric phase $\phi$ using $N$ particles~\cite{Giovannetti2004}.  A prerequisite for reaching the Heisenberg-limited uncertainty in a real measurement is a particle detector with atom number variance $\avar \ll 1$, i.e. exact particle counting at the interferometer output.  This capability is challenging to realize, particularly for large particle numbers.  For example, single-photon detectors suffer from limited quantum efficiency (typically $<$~95~$\%$), which prohibits resolving photon numbers for $N\gg10$~\cite{Hadfield2009}.  On the other hand, single atoms can be detected with near-unit efficiency by trapping them and observing their fluorescence~\cite{Neuhauser1980}.  Here, we extend this single-atom counting capability to mesoscopic atom numbers by high accuracy fluorescence measurements.

One example where single-atom resolution becomes necessary is spectroscopy with maximally entangled states.  Here, it has been shown that Heisenberg-limited precision requires measurement of the parity~\cite{Bollinger:1996uv}.  Another example is interferometry with spin-squeezed atomic states~\cite{Kitagawa1993}, where experimental results have shown a reduction of atom number variance approaching a level at which single-atom resolution becomes relevant~\cite{Ma2011}. Similarly, such high resolution atom detection would allow the direct observation of twin atom pairs produced via spin-changing collisions~\cite{Klempt2010, Gross2011, Hamley2012} and enable their use for interferometry at the Heisenberg limit.

\begin{figure}[hb!]
\includegraphics[width=0.45\textwidth]{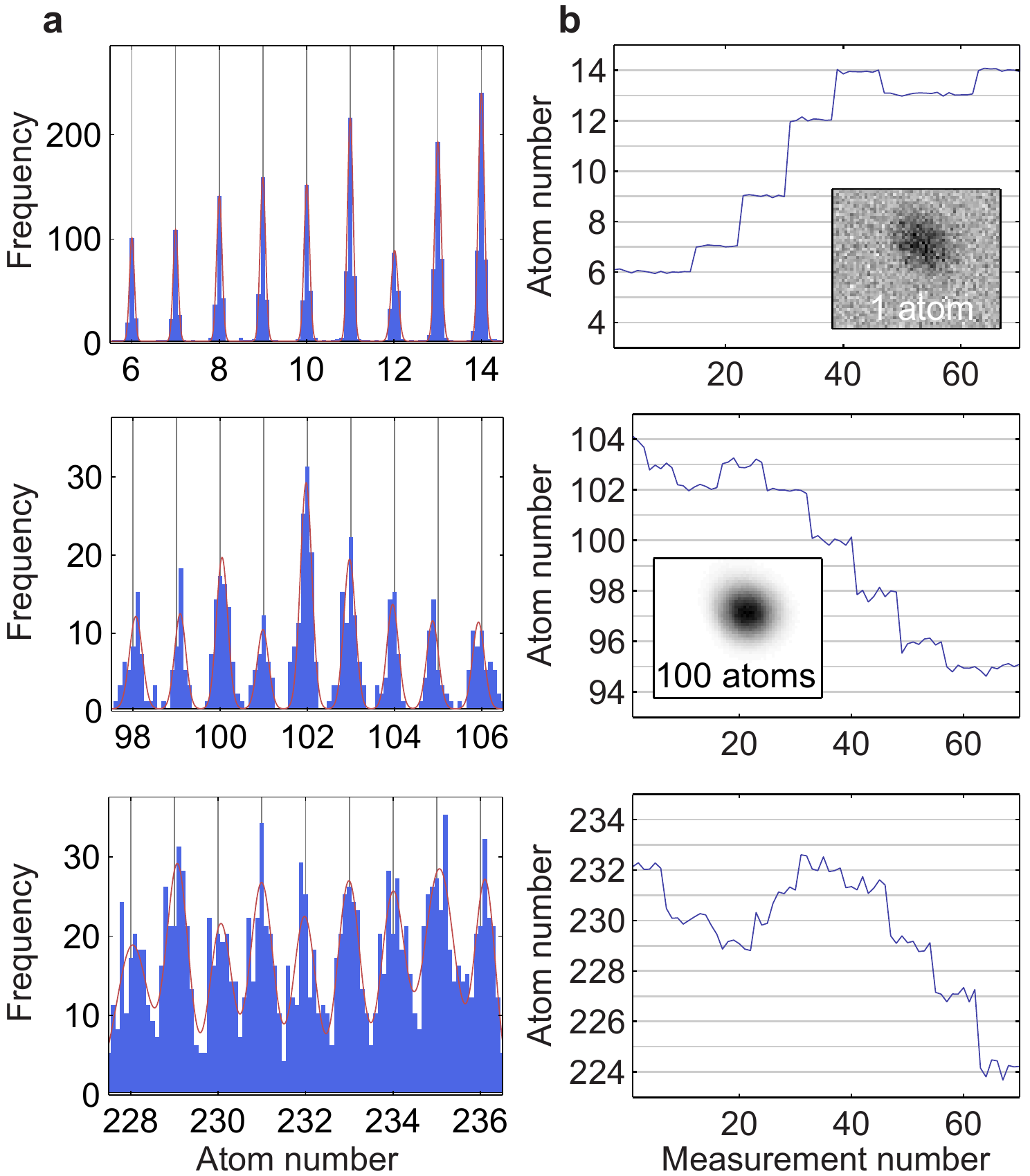}
\caption{\label{fig:Figure1}(a) Histogram of collected fluorescence signal (detection time $\td = 100$~ms) and Gaussian fits to the resulting distributions (red lines). (b) Example time traces for different fluorescence levels. The upper and center insets show the signal of a single atom and one hundred atoms, respectively.}
\end{figure}

The most common detection method for neutral atoms is absorption imaging, but the precision of such measurements on mesoscopic ensembles has thus far been limited to the level of a few atoms~\cite{Ockeloen2010, Muessel2013}.  Single-atom resolution for small atom numbers ($\NA\sim10$) has been achieved by fluorescence detection of neutral atoms in free space~\cite{Bucker:2009kl} as well as in magneto-optical traps (MOTs)~\cite{Hu:1994vq, Alt:2003il, Serwane2011}, optical dipole traps~\cite{Grunzweig:2010dr} and optical cavities~\cite{Puppe2007, Gehr2010, Goldwin2011}.  A recent experiment explored the detection of mesoscopic ensembles of atoms~\cite{Zhang:2012gw} in an optical cavity, and stability at the single-atom level was observed in repeated measurements for effective atom numbers as high as $\NA=150$.  In this case, however, accurate determination of the absolute atom number was not possible due to inhomogeneous coupling to the standing-wave probe light.  On the other hand, spatially resolved fluorescence measurements of atoms in optical lattices, can determine the number of singly-occupied sites~\cite{Nelson:2007ks, Bakr2010, Sherson:2010hg}, but atom pairs are quickly lost due to light-assisted collisions.  In contrast, here, we show exact counting of the total atom number in mesoscopic ensembles by fluorescence measurements in a MOT (Fig.~\ref{fig:Figure1}).

To estimate the limits of this approach, consider a fluorescence measurement of $\NA$ trapped atoms.  Two competing noise sources, fluorescence noise and noise from atom loss, determine the maximum atom number $\Nmax$ for which single-atom resolution is possible.  Photon-shot-noise (PSN) contributes a variance of $N/\np$ in terms of atom number when detecting $\np$ photons per atom in the absence of background photons.  The signal per atom can be expressed by ${\np = \Rsc\eta \td}$, where $\Rsc$ is the photon scattering rate, $\eta$ is the overall photon detection efficiency, and $\td$ is the detection time. The second noise source, atom number fluctuations due to trap loss, contributes a variance of $\NA \td/2\tau$, where $\tau$ is the trap lifetime. We can determine the optimal $\td$ by minimizing the total variance, ${\sigma^2_N  = N/\eta R_{\text{sc}} t + N t/2\tau}$.  Here we find $\topt = \sqrt{2 \tau/\eta \Rsc}$, which is independent of the atom number.  Furthermore, by setting $\sigN = 1$ we calculate an upper bound ${\Nmax = \sqrt{\tau \eta \Rsc/2}}$.  As a concrete example, consider $^{87}$Rb ($\Gamma/2\pi = 6.1$~MHz) trapped in a MOT.  Each atom fluoresces at a rate ${ R_{\text{sc}} = \Gamma/2\times s_0/(1+s_0+4\Delta^2/\Gamma^2)}$, where $s_0$ is the saturation parameter and $\Delta$ is the detuning of the laser from resonance.  If we assume typical experimental parameters (${s_0 = 1}$, $\Delta = -\Gamma/2$, $\eta = 0.01$ and $\tau = 100$~s), we find ${t_{\text{opt}}  = 56}$~ms and $\Nmax = 1800$.  This number is at least two orders of magnitude higher than the atom numbers counted in previous neutral-atom experiments.  In what follows we show measurements approaching this limit.

In our experiment, we image a MOT of ${\rm ^{87}Rb}$ atoms onto a low-noise CCD camera.  We estimate the total efficiency of the imaging system to be $\eta = 0.01$, which includes the numerical aperture of the aspherical objective lens (NA = 0.23), the camera quantum efficiency, and the transmission of all optical elements.  The MOT beam diameters typically have a waist of $w=1.5$~mm during atom counting and the peak intensity, summed over all six beams, is 23~mW/cm$^2$, corresponding to a saturation parameter $s_0 \simeq 6.5$. From this, we estimate the scattering rate per atom as ${R_{\text{sc}} = 15\times10^{6}}$~s$^{-1}$ at the detuning of approximately $-\Gamma/2$.

The histograms in Fig.~\ref{fig:Figure1} are generated by binning repeated fluorescence measurements over an 8 hour time period.  The effective detection time for the measurements is 100 ms, where each measurement integrates the fluorescence from two adjacent 50 ms exposures.  The background count level, recalibrated every 15 minutes, is typically less than the signal from 3 atoms.  For atom numbers as high as $N \sim 300$, resolved peaks appear in the fluorescence histograms corresponding to the signal from an exact number of atoms.  Over the same range of atom numbers, steps can be observed in the time-resolved fluorescence signal, coinciding with the loading or loss of individual atoms.  These features indicate atom number resolution significantly below the single-atom level.

Based on the resolved histogram peaks, we can characterize several properties of the detector.  First, by fitting the peaks to a sum of equally-spaced Gaussian distributions, we calibrate the single-atom count rate to be $90310$~counts/atom/s.  A quadratic fit to the centroid of all resolvable peaks as a function of $\NA$ reveals no evidence for nonlinear scaling of the count rate with atom number.   The uncertainty in the second-order fit coefficient is one way to quantify the calibration accuracy, and we constrain the deviation from linearity to below 0.02~\% at $\NA = 250$ (95 \% confidence interval).  The width of the individual distributions are a measure of fluorescence noise for a given atom number.  For example, at $\NA=100$ we find a standard deviation of $\sigma = 0.14$~atoms growing to $\sigma = 0.27$~atoms at $\NA = 230$.  These numbers are comparable to the expected photon shot noise of $\sigma_{\text{psn}} = 0.11$ at $\NA = 100$ and $\sigma_{\text{psn}} = 0.16$ at $\NA = 230$.

\begin{figure}[tb]
\includegraphics[width=0.45\textwidth]{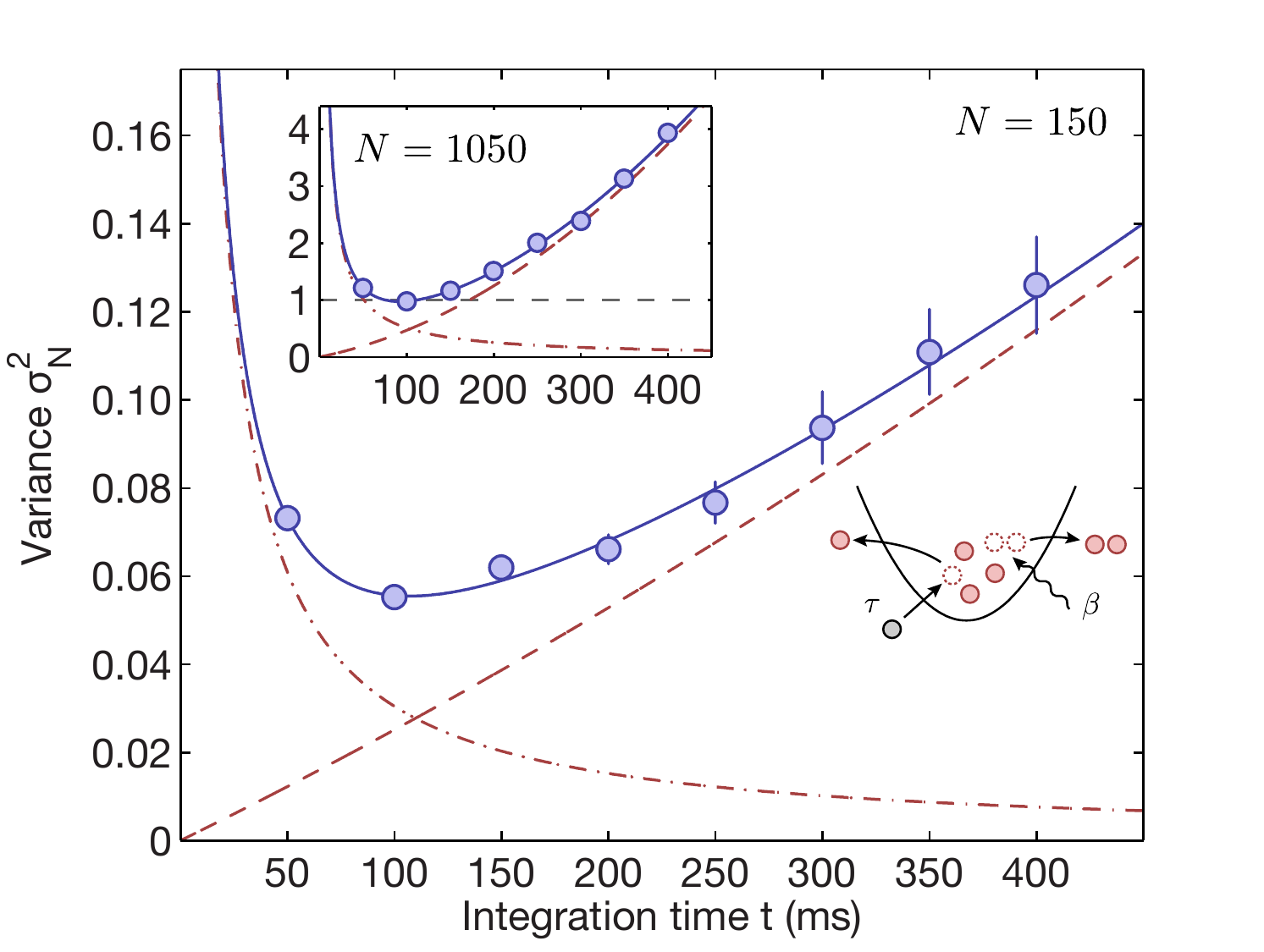}
\caption{\label{fig:Figure2} Atom number variance as a function of exposure time.  For atom numbers in the range $\NA = 100$ to $200$, the measured variance (circles) reaches a minimum near ${t = 100}$~ms.  For short integration times $\avarA$ is limited by  photon shot noise and additional fluorescence noise, which average in time (dash-dot line). For long integration times, the finite lifetime is the dominant noise contribution (dashed line), where the main loss processes, depicted in the diagram, are collisions with background gas and light-assisted collisions.  Higher atom numbers ($\NA = 1000$ to $1100$ in the inset) exhibit a higher overall variance, but the optimal detection time is found to be similar.  The error bars represent 1-$\sigma$ statistical uncertainties and the solid lines are fits based on the model described in the text.}
\end{figure}

To characterize the detection noise $\sigN$ in a general way for higher atom numbers, we calculate the two-sample atom variance (equivalent to the Allan variance in frequency measurements)
\begin{equation}
\label{eq:avar}
\avarA = \frac{1}{2} \left< (S_{n+1} - S_{n})^2 \right>,
\end{equation}
where $S_{n}$ and $S_{n+1}$ are the signals of consecutive measurements, each integrating CCD counts for time $\td$. This measurement captures both fluorescence noise and number fluctuations due to atom loss, but, in contrast to the histograms above, is not susceptible to long term drifts in the signal.  Fig.~\ref{fig:Figure2} shows the results of such an analysis for atom numbers in the range $\NA = 100$ to $200$.  It indicates that there is an optimal detection time, ${\topt\sim 100}$~ms, after which atom loss begins to dominate the noise.  The same analysis for $\NA = 1000$ to $1100$ (Fig.~\ref{fig:Figure2},~inset) shows that $\topt$ does not change significantly over the full range of atom numbers, as expected.

\begin{figure}[tb]
\includegraphics[width=0.5\textwidth]{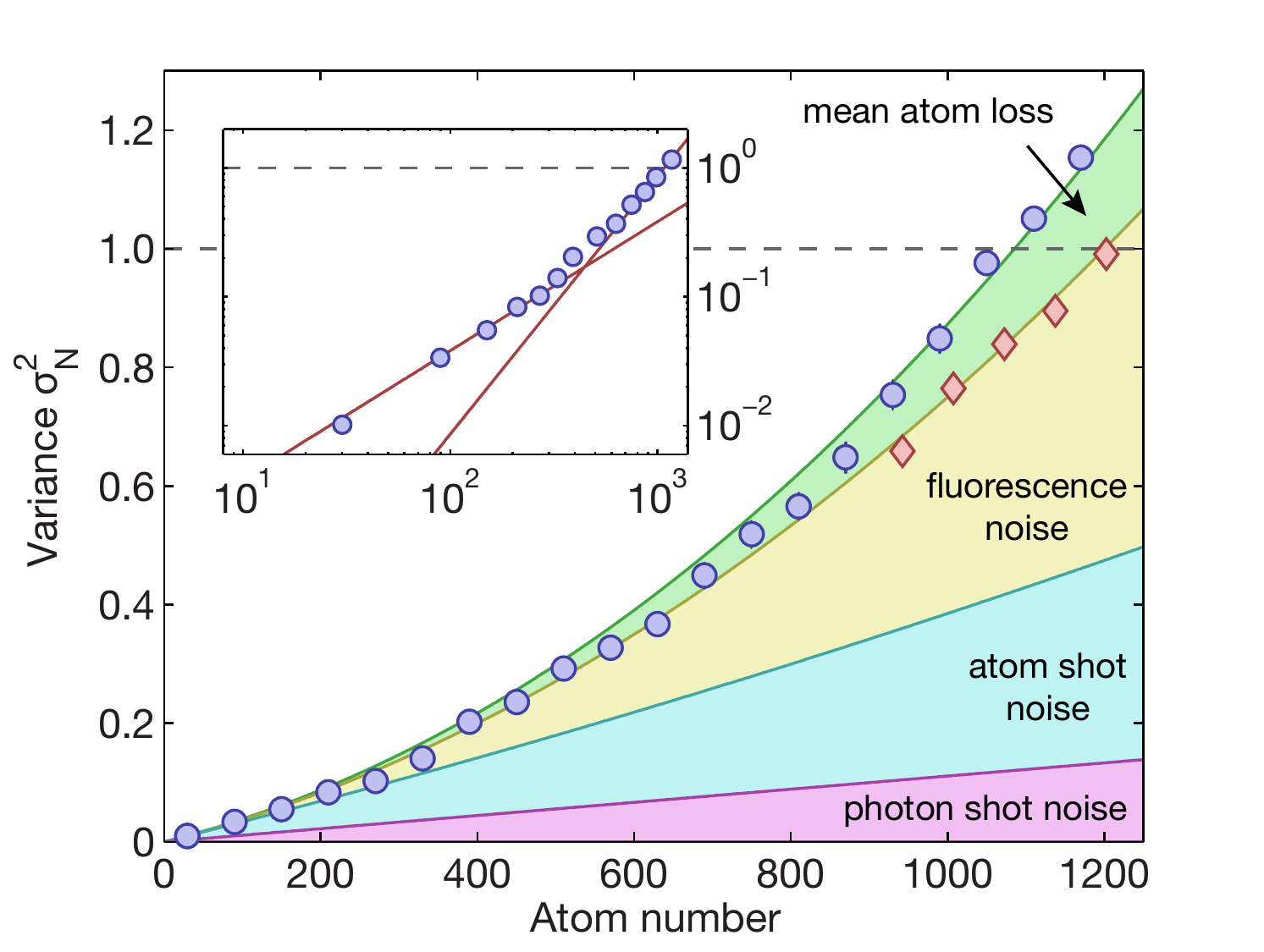}
\caption{\label{fig:Figure3} Single-atom resolution for mesoscopic atom numbers. For the near-optimal exposure time of $\SI{100}{\milli\second}$ the variance $\avarA$ reaches the limit of single-atom resolution at ${N = 1080}$. A fit to the data based on a noise model incorporating fluorescence noise and atom loss (upper green line) allow us to estimate the relative contribution of these noise sources (shaded regions) and compensate for the mean atom loss (red diamonds) as described in the text.  A logarithmic plot (inset) shows how the scaling of the variance with respect to atom number transitions from linear to quadratic as the additional fluorescence noise begins to dominate the shot noise.}
\end{figure}

Taking the 100 ms detection time, we determine the variance as a function of atom number up to $\NA = 1200$ (Fig.~\ref{fig:Figure3}). Here, it can be seen that the single-atom resolution threshold, $\sigNA = 1$, is reached near $\NA = 1080$.  Viewing the same data on a logarithmic plot (Fig.~\ref{fig:Figure3}, inset) shows how the variance initially scales with $\NA$ at low atom numbers then changes to scaling with $\NA^2$ at higher atom numbers.

To better understand the limiting noise sources we fit an equation of the following form to the data,
\begin{equation}
\label{eq:SigN}
\avarA = a(\NA)t^{-1}+b(\NA)t+c(\NA)t^2,
\end{equation}
where the first term represents fluorescence noise and PSN, while the last two terms come from atom loss -- both the atom shot noise due to discrete loss events, quantified by $b(\NA)$, and the decay of the mean number of atoms, quantified by $c(\NA)$.  In particular, we use ${a(\NA) = \frac{\NA}{\eta \Rsc}+(\sigf\NA)^2}$, where the first term represents photon shot noise and the second term is an additional noise source that is assumed to be uncorrelated in time but common to all atoms in the MOT.  This describes, for example, fast frequency or intensity noise on the MOT laser beams.  To determine $b(\NA)$ and $c(\NA)$ we use a master equation approach based on the rate equation ${dN/dt = -N/\tau - \beta N^2}$, where both one-body loss, parameterized by $\tau$, and two-body loss, parameterized by $\beta$, are considered (see supplementary information~\cite{Supplement}).  The resulting noise coefficiencts are given by ${b(\NA) = \NA/2\tau + \beta \NA^2}$, which is the dominant effect of loss in our measurements, and ${c(\NA) = \left(\NA/\tau + \beta \NA^2\right)^2/2}$.

A fit of the noise model to the data, for which we vary $\sigf$, $\tau$ and $\beta$, is performed individually for each atom number.  Two examples of such fits are shown in Fig.~\ref{fig:Figure2}.  We take the means of the independent fit parameters as inputs to the noise model to produce the curve in Fig.~\ref{fig:Figure3}.  We extract a value for the fluorescence noise parameter of ${\alpha=1.9(1)\times 10^{-4}}$~s$^{1/2}$ (uncertainties represent 1-$\sigma$ statistical uncertainty).  A likely source of this additional noise in our experiment is frequency noise on the MOT lasers, which would correspond to about 10~kHz deviations in detuning when averaged over the detection time. We also extract the loss parameters $\tau = 246(44)$~s and $\beta = 3(3)\times10^{-7}$~s$^{-1}$, indicating that light-assisted collisions contribute only  a small amount to the detection errors at these atom numbers.  Since the atom loss is well known for all atom numbers based on the calibrated parameters, we can improve the measurement accuracy by compensating for the loss that occurs during detection.  If the raw measurement yields the result $\NA$, then one computes a loss-compensated result $\NA^{\prime} = \NA + \NA\td/2\tau$, neglecting $\beta$. Assuming proper measurement calibration, the limiting noise is then given by $\text{Var}(S_{n+1} - S_{n})/2$.  Computing this variance for the same data set yields a threshold for single atom resolution of $\NA = 1200$, coinciding with the sum of remaining noise terms, as shown in Fig.~\ref{fig:Figure3}.  To illustrate the meaning of these variances in terms of atom counting, consider the fidelity, here defined as the probability of exactly identifying the initial atom number. We evaluate this based on a Monte-Carlo simulation assuming the measured count rate, fluorescence noise and loss parameters and find a fidelity of 99.8~\% at $\NA = 10$,  98.5~\% at $\NA = 100$, and 44~\% at $\NA = 1200$ (details in supplementary information~\cite{Supplement}).

\begin{figure}[t]
\includegraphics[width=0.4\textwidth]{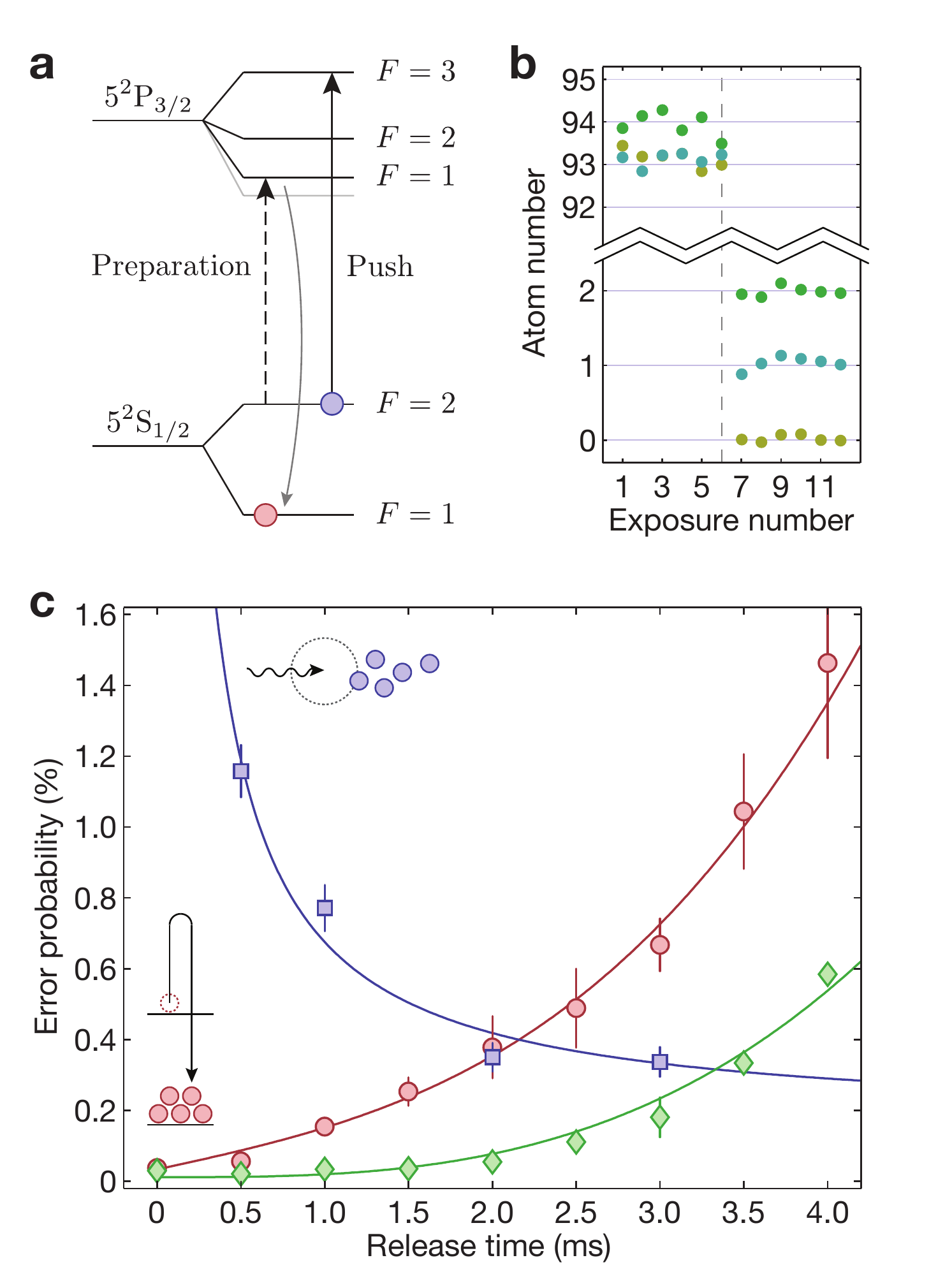}
\caption{\label{fig:Figure4} Efficiency of state-selectivite detection via radiation pressure. (a) Level scheme of the $^{87}$Rb D$_2$ transition. The push beam accelerates atoms in $|F = 2\rangle$ out of the capture volume while the MOT beams and repumper are off.  The atoms can be prepared in $|F = 1\rangle$ via optical pumping. (b) Three example measurements of the $\ket{F=2}$ push efficiency, detecting the atom number before and after the push pulse (dashed line).  (c) The measured error rate without push pulse (diamonds), is consistent with isotropic thermal expansion (green line).  The errors for $\ket{F=2}$ atoms (squares) can be qualitatively understood based on an analysis of the rate of depumping from $|F = 2\rangle$ to $|F = 1\rangle$ including the effects of imperfect push beam polarization, and imperfect state preparation (blue line).  The error rate for atoms prepared in $\ket{F=1}$ (circles) includes both the effects of thermal expansion and off-resonant optical pumping (red line).  All error bars represent 1-$\sigma$ statistical uncertainties.}
\end{figure}

We now investigate state-selective detection of the two hyperfine levels, $\ket{F=1}$ and $\ket{F=2}$, in the $^2$S$_{1/2}$ manifold (Fig.~\ref{fig:Figure4}).  The technique is based on release and recapture of the atoms, where, during the release, atoms in $\ket{F=2}$ are pushed out of the capture volume by resonant radiation pressure. As a starting point, we measure the efficiency of the release and recapture, as a function of the release time, without radiation pressure. By counting the atoms before and after the release, we find a recapture fidelity above 99.92 \% for release times up to 2 ms.  To distinguish the populations we apply a laser pulse resonant with the ${|F = 2\rangle\rightarrow \rm{|^2P_{3/2}}, F=3\rangle}$ transition during the release time (between two exposures in a single CCD frame), which imparts momentum to the $\ket{F=2}$ atoms, while ideally leaving the $\ket{F=1}$ atoms unaffected.  We measure the overall error probability for the two cases when the atoms are prepared in either $\ket{F=2}$ or $\ket{F=1}$ via optical pumping.  The release time with equal error probability for both states is found to be $\SI{2.2}{\milli\second}$,  where we measure an average fidelity of 99.6(1)~\%, sufficient for detecting the state of 250 atoms with single-atom resolution.

In summary, we have shown single-atom resolution for fluorescence measurements of up to 1200 atoms.  Our results show that a MOT, with high trap depths, low atom density and high photon scattering rate, is a near ideal trap for precise fluorescence measurements.  Since the basic experimental techniques used here are common to many cold-atom experiments, an atom detector with this level of performance could be implemented in many contexts.  In addition, our noise analysis is relevant for fluorescence measurements in other atom traps such as optical dipole traps, where we expect both atom loss from light-assisted collisions and photon shot noise to be more severe constraints.  In the present work, we demonstrated state-selective detection using radiation pressure to separate two hyperfine states, counting the atoms of one state that remain in the trap volume.   However, for many experiments it would be advantageous to simultaneously detect the atom number in two or more sub-ensembles.  For this, we envision a system where atoms in the relevant quantum states are separated spatially then individually trapped and detected via fluorescence measurements.  We are currently developing this capability in our experiment.  When combined, for example, with Stern-Gerlach separation, this will enable measurements of spin-entangled BECs at the Heisenberg limit.

\begin{acknowledgments}
The authors acknowledge experimental contributions by Patrick Reiser.  This work was supported by the Heidelberg Center for Quantum Dynamics, Forschergruppe FOR760 of the Deutsche Forschungsgemeinschaft, and the FET-Open project QIBEC (Contract No. 284584).  D. H. acknowledges support from the Alexander von Humboldt Foundation, and I. S. from the International Max Planck Research School (IMPRS-QD).
\end{acknowledgments}

\pagebreak
\clearpage

\section{Supplementary Material}
\setcounter{figure}{0}
\setcounter{equation}{0}
\setcounter{page}{0}
\renewcommand{\thepage}{S\arabic{page}}
\renewcommand{\theequation}{S\arabic{equation}}
\renewcommand{\thesection}{S\arabic{section}}
\renewcommand{\thetable}{S\arabic{table}}
\renewcommand{\thefigure}{S\arabic{figure}}
\renewcommand{\bibnumfmt}[1]{[S#1]}
\renewcommand{\citenumfont}[1]{\textit{S#1}}

\subsection{Additional Experimental Details}

Our experiment requires maintaining a stable MOT with a long lifetime, low loading rate from background gas, and low levels of stray light from the MOT and repumper beam.  We have taken several steps to ensure these conditions.  First, the size of the MOT beams is adjusted by a motorized iris between a waist of 17 mm and 1 mm.  During detection the MOT beam waist is typically 1.5 mm and we achieve a background count rate below 225~kcts/s, or roughly equivalent to the signal from 2.5 atoms.  Similarly, we use a single repumping beam centered on the MOT with a waist below 1 mm to minimize the total incident power and resulting stray light.  This makes the contribution of background photons to our detection noise negligible for almost all atom numbers considered.  The reduced beam size also limits the loading rate of atoms from background gas to levels of $4\times10^{-3}$~1/s and below, so that, for most atom numbers the loading rate is insignificant compared to the loss.  Interference fringes on the MOT beams, particularly due to diffraction through the iris, can lead to inhomogenous intensity imbalances across the MOT, and result in fluorescence noise and heating~\cite{Chaneliere2005}.  For this reason, after the motorized iris, we implement a spatial filter that is optimized for the small beam sizes, where these effects are most important.

The image of the MOT is masked spatially in an intermediate image plane, so that it illuminates only a narrow stripe of the CCD.  This has also been found to reduce stray light, by providing some level of spatial filtering.  Using a subframe transfer mode of the camera, we make a series of measurements in rapid succession before a mechanical shutter is closed and the frame is read out (3.5~s readout time), so that the timing between adjacent exposures is controlled at the sub-millisecond level.  The CCD read noise (4 e$^-$/pixel) is negligible for all measurements.

\subsection{Noise Model and Master Equation}
In order to understand the limiting noise contributions for atom number detection and find the optimal exposure time, we evaluate the atom variance $\sigma_N^2$ for different atom numbers and integration times. At short integration times, we expect the major noise contribution to arise from fluorescence noise, consisting of photon shot noise and scattering rate noise, due to frequency or intensity fluctuations of the excitation light. For longer integrations times, the primary noise source is the lifetime of atoms in the trap, due to one- and two-body decay. The general model is
\begin{equation}
\label{eq:NoiseModel}
\sigma^2_{N} = \sigma^2_{\mathrm{psn}} + \sigma^2_{\mathrm{srn}} + \sigma^2_{\mathrm{loss}}.
\end{equation}
Photon shot noise is given by $\sigma^2_{\mathrm{psn}} = \eta \Rsc N t^{-1}$, where $(\eta \Rsc)^{-1}$ is the count rate per atom. Scattering rate noise is described by $\sigma^2_{\mathrm{srn}} = (\alpha N)^2 t^{-1}$, which arises from correlated fluctuations in the scattering rate for each of the $N$ atoms.

The two primary processes that lead to atom loss in a magneto-optical trap are one-body decay due to collisions with the background gas, parameterized by the lifetime $\tau$, and light-assisted collisions, where two atoms are lost simultaneously, parameterized by $\beta$. Because of the low density in a MOT, three-body losses can be neglected.  The rate equation for atom loss is given by,
\begin{equation}
\label{eq:dNdt}
\frac{dN}{dt} = -\frac{N}{\tau} - \beta N^2.
\end{equation}
We are interested in the effects of loss over the detection time $t$, which is short compared to $\tau$ and $\beta N$ for all $N$ considered, so that the relevant part of the loss curve is just the initial linear component,
\begin{equation}
\label{eq:Nt}
N(t) = N_0-\left(\frac{N_0}{\tau} + \beta N_0^2\right) t.
\end{equation}
The noise $\sigma^2_{\mathrm{loss}}$ due to atom loss depends on the nature of the loss process, i.e. one-body and two-body processes will contribute differently.  These contributions can be derived from a master equation~\cite{VanKampen2007}.

We consider the probability $P(N,t)$ of having $N$ atoms in the trap at a given time. The time evolution of this probability is given by the master equation
\begin{equation}
\label{eq:MasterEq}
\displaystyle
\frac{\partial}{\partial t} P(N,t) = \sum_{N^{\prime}} W_{N^{\prime},N} P(N^{\prime},t) - W_{N,N^{\prime}} P(N,t).
\end{equation}
The first term describes an increase of the probability $P(N,t)$ when the atom number changes from any $N'$ to $N$ with a weight $W_{N^{\prime},N}$. The second term accounts for a change in atom number from $N$ to any $N^{\prime}$, weighted with $W_{N,N^{\prime}}$, which reduces the probability.  The master equation for one- and two-body decay reads
\begin{widetext}
\begin{equation}
\label{eq:dPdt}
\frac{\partial}{\partial t} P(N,t) = \frac1{\tau} \left[(N+1) P(N+1,t) - N P(N,t)\right] + \frac{\beta}{2} \left[(N+2)(N+1) P(N+2,t) - N(N-1) P(N,t)\right],
\end{equation}
\end{widetext}
where $\beta$ has to be halved, since by definition it accounts for the loss of only one atom, while in the actual process two atoms are lost from the trap.  The time evolution of the $k$-th moment of $P(N,t)$ is given by
\begin{eqnarray}
\label{eq:dNkdt}
\displaystyle
\frac{\partial}{\partial t} \left<N^k(t)\right> & = & \frac{\partial}{\partial t} \sum_N N^k P(N,t)\nonumber\\
& = & \sum_N N^k \frac{\partial}{\partial t}P(N,t).
\end{eqnarray}
Eq. \ref{eq:dPdt} can be substituted in Eq. \ref{eq:dNkdt} to generate a set of coupled differential equations for all moments of $P(N,t)$.  We are interested in the evolution of the variance ${\text{Var}\left(N\right) = \left<N^2\right> - \left<N\right>^2}$.  To evaluate the differential equations we apply two conditions that are easily satisfied in our experiments.  First, ${\text{Var}\left(N\right) \ll \left<N\right>}$, as required for a precise atom number measurement.  Second, ${\left<N\right>^2 \gg \left<N\right>}$, which is true for the mesoscopic atom numbers of interest.  With these, we obtain
\begin{equation}
\label{eq:sigmaN}
\frac{\partial}{\partial t} \text{Var}(N) = \frac{\left<N\right>}{\tau} + 2 \beta \left<N\right>^2.
\end{equation}
For $t_d \ll \tau$ and $t_d\ll \beta N_0$ as in our experiments,
\begin{equation}
\label{eq:VarN}
\text{Var}\left(N\right) = \frac{N_0 t}{\tau} + 2 \beta N_0^2 t.
\end{equation}

Experimentally, we measure the two-sample variance ${\sigma^2_N = \frac{1}{2} \left< (S_{n+1} - S_n)^2 \right>}$, where $S_{n+1}$ and $S_{n}$ are the consecutive detected atom numbers.  Considering only loss, this can be written,
\begin{eqnarray}
\sigma^2_{\mathrm{loss}} & = & \frac{1}{2} \left< (N(t) - N_0)^2 \right>\nonumber\\
& = & \frac12 \text{Var}(N(t)) +  \frac{1}{2}\left<N(t)-N_0\right>^2.
\end{eqnarray}
Using Eq.~\ref{eq:Nt} and Eq.\ref{eq:VarN}, we have
\begin{equation}
\sigma^2_{\mathrm{loss}} = \frac{N_0}{2 \tau} t + \beta N_0^2 t + \frac{1}{2} \left( \frac{N_0}{\tau} + \beta N_0^2 \right)^2 t^2,
\end{equation}
which is the last term in the full noise model Eq.~\ref{eq:NoiseModel}.

\subsection{Monte Carlo Simulation for Fidelity Estimates}

\begin{figure}[b]
\includegraphics[width=0.45\textwidth]{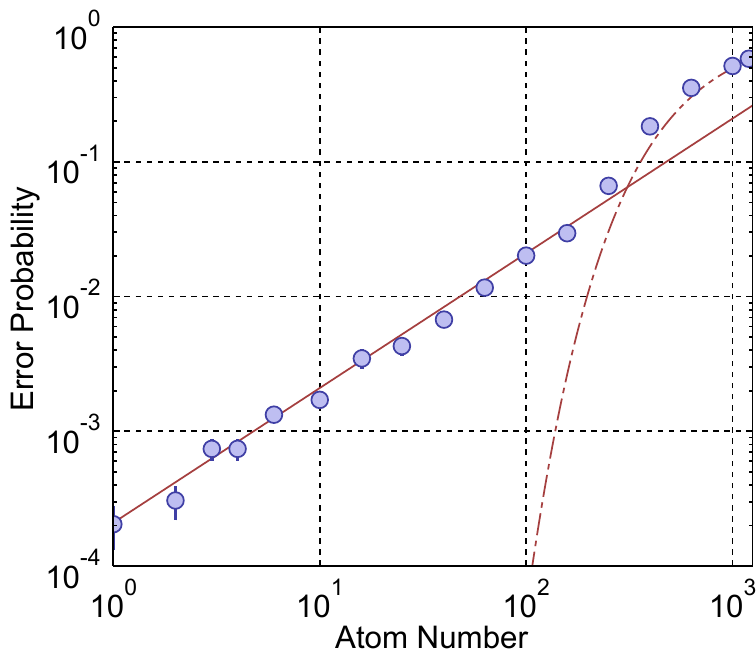}
\caption{\label{fig:Fidelity} Error probability for determining the exact atom number.  Results from a Monte-Carlo simulation (circles) show how the error probability $1-F$ increases with atom number.  At low atom numbers the loss probability $P_{\mathrm{loss}} \simeq N_0 \td/2\tau$ (solid line) dominates the errors.  At higher atom numbers, fluorescence noise (dash-dotted line) is the primary source of error. }
\end{figure}

The variance of our fluorescence measurements as a function of atom number as shown in Fig. 3 (main text) specifies the range over which we achieve single-atom resolution ($\sigN < 1$).  In the context of measuring a discrete quantity such as the atom number an intuitive figure of merit is the fidelity $F$, which we define as the probability that the measurement outcome corresponds to the exact number of atoms in the ensemble at the beginning of the measurement.  Given the initial atom number, the effects of atom loss lead to a non-Gaussian distribution of measurement outcomes, such that the fidelity cannot be directly calculated from the measurement variance alone.  For this reason, to estimate this fidelity, we take the noise parameters, $\tau$, $\beta$, and $\alpha$, extracted from the experimental measurements and perform a Monte-Carlo simulation of the measurement process with a known initial atom number, $N_0$.  This allows us to estimate the fidelity $F$, or equivalently the error probability $1-F$ (Fig.~\ref{fig:Fidelity}).

Beginning from $N_0$, the simulation determines the atom number $N_k$, and the number of CCD counts on a time interval $\dt$, which is chosen to be much less than the total exposure time $t=100$~ms.  At each time step, the number of atoms lost is randomly chosen from a convolution of distributions that quantify one-body and two-body loss.  These are the same probabilities entering the master equation, Eq.~\ref{eq:dPdt}, above.  The dominant contribution for most atom numbers is the probability of losing a single atom due to collisions with background gas, $P_1 \simeq N_k \dt/\tau$.  For all atom numbers we consider, this probability is less than $1$~\% over the simulation time interval $\dt$.  We also take into account two-body loss events resulting in the loss of both atoms, with probability $P_2 \simeq \beta N^2_k \dt/2$.

Given the $N_k$ trapped atoms during $\dt$ we randomly generate a number of CCD counts $c_k$ based on a Gaussian distribution with mean $\cmean_k = N_k \eta \Rsc \dt$ and variance ${\sigc^2 = \left(\sigma^2_{\mathrm{psn}} + \sigma^2_{\mathrm{srn}}\right)\eta^2 \Rsc^2}$, taken from the noise model described above and evaluated at $\dt$.  The total counts $C=\sum_k c_k$ summed over all k within the exposure time is used to determine the measurement result $N_{\mathrm{meas}} = C/\td$.  Here we ignore counts from stray light, which contributes negligibly to the noise for all atom numbers.  We calculate a loss-compensated result ${N_{\mathrm{meas}}^{\prime} = N_{\mathrm{meas}}(1+\td/\tau+\beta N_{\mathrm{meas}}\td)}$, and the final detection outcome comes from rounding $N_{\mathrm{meas}}^{\prime}$ to the nearest whole number.

When we compare these outcomes to $N_0$, we find the error probabilities plotted in Fig.~\ref{fig:Fidelity}.  For atom numbers up to $N_0=200$ the probability of a one-body loss event, $P_{\mathrm{loss}} \simeq N_0 \td/2\tau,$ dominates these errors. Note that, in the absence of fluorescence noise, only loss events during the first half of the detection time will result in an error leading the the factor of 2 in $P_{\mathrm{loss}}$.  Above $N_0=200$ fluorescence noise becomes more significant.  This noise is described by a normal distribution ${f(N) = e^{-(N-N_0)^2/2\sigma_{\mathrm{f}}}/\sqrt{2}\sigma_{\mathrm{f}}}$, where $\sigma^2_{\mathrm{f}} =\sigma^2_{\mathrm{psn}} + \sigma^2_{\mathrm{srn}}$ from Eq.~\ref{eq:NoiseModel} depends on $N_0$.  The probability that a measurement lies outside the thresholds at $N_0-\frac12$ and $N_0+\frac12$ is,
\begin{equation}
P_{\mathrm{f}} = 1+\mathrm{erf}\left(-\frac{1}{\sqrt{8}\sigma_{\mathrm{f}}}\right).
\end{equation}
Both $P_{\mathrm{f}}$ (dash-dotted line) and $P_{\mathrm{loss}}$ (solid line) are plotted along with the simulation data in Fig.~\ref{fig:Fidelity}, and show excellent agreement with the simulation in the appropriate atom number ranges.

%



\end{document}